\newcommand{\md}{\mathrm{d}}

\newcommand{\be}{\begin{equation}}
\newcommand{\ee}{\end{equation}}
\newcommand{\bea}{\begin{eqnarray}}
\newcommand{\eea}{\end{eqnarray}}
\newcommand{\bean}{\begin{eqnarray*}}
\newcommand{\eean}{\end{eqnarray*}}

\newcommand{\Jpsi}{J/\psi}

\documentclass[12pt]{iopart}

\usepackage{iopams} 

\usepackage{epsfig,latexsym,amssymb,cite}
\usepackage{graphicx}
\usepackage{bm}

\begin{document}

\title{Heavy Quark diffusion from lattice QCD spectral functions}

\author{H.-T. Ding$^a$, A. Francis$^b$, O. Kaczmarek$^b$,  F. Karsch$^{a,b}$, H. Satz$^b$, W. S\"oldner$^{c}$}

\address{$^{\rm a}$ Physics Department, Brookhaven National Laboratory, Upton, NY 11973, USA \\
$^{\rm b}$ Fakult\"at f\"ur Physik, Universit\"at Bielefeld, D-33615 Bielefeld, Germany\\
$^{\rm c}$ Institut f\"ur Theoretische Physik, Universit\"at Regensburg, D-93040 Regensburg, Germany\\}

\ead{htding@bnl.gov}
\begin{abstract}
	We analyze the low frequency part of charmonium spectral functions on large lattices close to 
the continuum limit in the temperature region $1.5\lesssim T/T_c\lesssim 3$ as well as for $T \simeq 0.75T_c$. 
We present evidence for the existence of a transport peak above $T_c$ and its absence below $T_c$. 
The heavy quark diffusion constant is then estimated using the Kubo formula. As part of the calculation 
we also determine the temperature dependence of the signature for the charmonium bound state in 
the spectral function and discuss the fate of charmonium states in the hot medium.

\end{abstract}

\section{Introduction}

Experimentally a substantial elliptic follow of heavy quarks has been observed at RHIC~\cite{Adare:2006nq}. 
Various phenomenogical model studies suggest heavy quark diffusion coefficient $D\lesssim1/T$ to accomodate data~(see e.g. Ref.\cite{Moore:2004tg}).
Theoretically the heavy quark diffusion coefficient $D$ has been calculated by perturbative QCD 
in both leading and next-to-leading order as well as from AdS/CFT correspondence. At $\alpha_s\approx 0.2$, 
leading order pQCD calculation gives $2\pi TD\approx 71.2$~\cite{Moore:2004tg}  while next-to-leading 
order calculation gives $2\pi TD\approx 8.4$~\cite{CaronHuot:2007gq}. In the strong coupling limit 
$2\pi TD=1$ is obtained from AdS/CFT correspondence~\cite{Kovtun:2003wp}. Under such a circumstance 
a non-pertubative computation of heavy quark diffusion is needed\footnote{Here we focus on extracting diffusion coefficient from mesonic spectral functions, 
other ways can be found in Ref.~\cite{CaronHuot:2009uh}.}.

 Through Kubo formula, the heavy quark diffusion constant $D$ relates to the vector spectral function as
$D= \frac{\pi}{3\chi_{00}}\lim_{\omega\rightarrow0}\sum_{i=1}^{3}\frac{\sigma_{V}^{ii}(\omega,T)}{\omega},
\label{HQ_diff}$
where $\chi_{00}$ is the quark number susceptibility and $\sigma_{V}^{ii}(\omega,T)$ is a mesonic 
spectral function in the vector channel. In the non-interacting case $\sigma_{V}^{ii}(\omega)$ has 
a $\omega\delta(\omega)$ term~\cite{Karsch:2003wy} and consequently gives an infinity diffusion constant, while in the interacting case 
$\omega\delta(\omega)$ will be smeared into a Breit-Wigner form~\cite{Petreczky:2005nh} and leads to a finite diffusion.
The fate of quarkonia states at finite temperature, which was suggested as a useful probe of QCD medium properties~\cite{Satz86}, 
can also be signaled by the deformation of spectral functions.

The mesonic spectral function is not directly accessible through lattice QCD simulations and can be obtained 
from the inversion of the following equation
\be
\hspace{-1cm}G_{H}(\tau,T)=\int_0^{\infty}{\mathrm{d}\omega~\sigma_{H}(\omega,T)}~K(\tau,T,\omega),
~~K(\tau,T,\omega)=\frac{ \mathrm{cosh}(\omega(\tau-\frac{1}{2T}))}{\mathrm{sinh}(\frac{\omega}{2T})}.
\label{eq:relation_cor_spf}
\ee
where the two-point correlation function $G_{H}(\tau,T)=\sum_{\vec{x}} \langle~ J_H(\tau,\vec{x})~J_{H}^{\dag}(0,\vec{0})~\rangle_T$ can be computed on the lattice. 
$J_H=\bar{q}(\tau,\vec{x})\Gamma_{H} q(\tau,\vec{x})$ is a local mesonic operator and $\Gamma_{H}=\gamma_{i},\gamma_{5}$ 
for vector ($V_{ii}$) and pseudo-scalar ($PS$) channels, respectively. 
The temperature $T$ is related to the Euclidean temporal extent $aN_{\tau}$ by $T=1/(aN_{\tau})$, where $a$ is the lattice spacing.
 We measured charmonium correlation functions on very fine ($a=0.01fm$) 
quenched lattices with a relatively large size of $128^{3}\times 96$, $128^3\times48$, $128^3\times32$ 
and $128^3\times24$ at $0.73~T_c$, $1.46~T_c$, $2.20~T_c$ and $2.93~T_c$, respectively. 
The lattice parameters and part of the results have been reported in Ref.~\cite{Ding:2010yz}.

\begin{figure}[b]
  \begin{center}
  \includegraphics[width=0.4\textwidth]{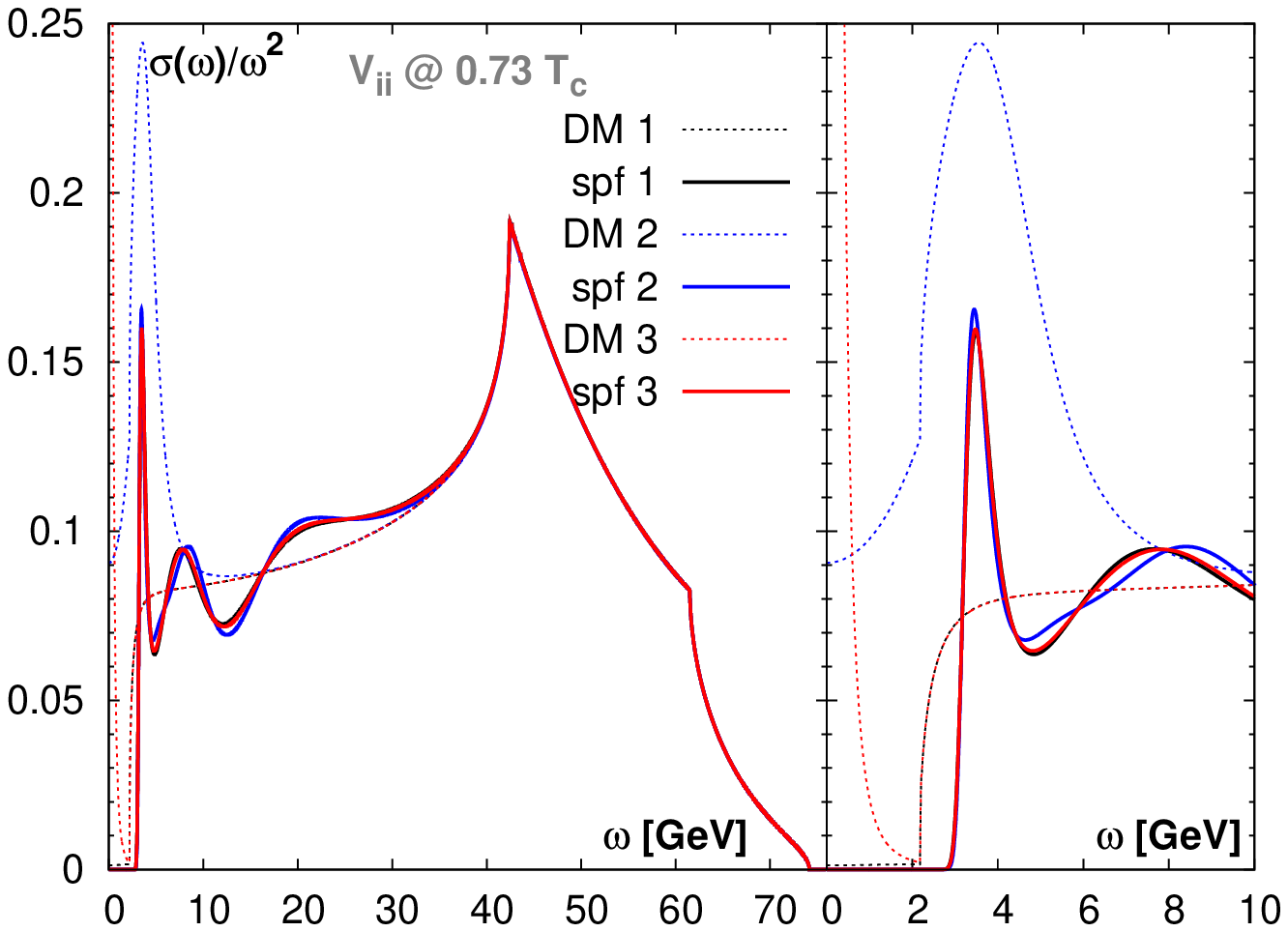}~
  \includegraphics[width=0.4\textwidth]{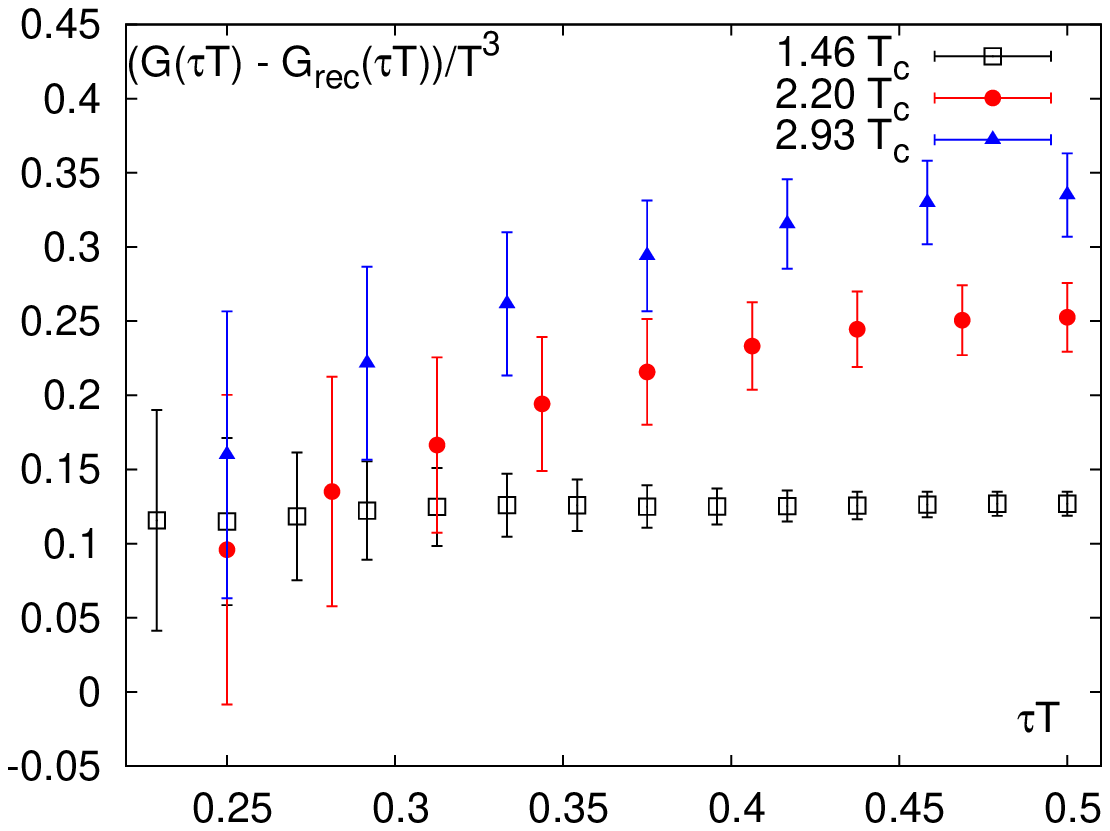}~
  \caption{Left: Default model dependences of vector spectral functions at $0.73~T_c$. 
           Right: differences of $G(\tau T)$ from $G_{rec}(\tau T)$ as a function of $\tau T$. $G_{rec}(\tau T)$ is the reconstructed 
           correlation function from the spectral function at $0.73 T_c$~\cite{Ding:2010yz}.
           }
                 \label{fig:cor}
\end{center}
\end{figure}

\section{Default model dependences of MEM results}
The Maximum Entropy Method (MEM) is used to extract spectral functions from correlators through 
Eq.~(\ref{eq:relation_cor_spf})~\cite{Asakawa01}. The input parameter of MEM is a default model (DM), which includes
the knowledge about the spectral function, e.g. $\sigma(\omega)\ge0$. The output spectral functions from
MEM are reliable only if they show a small dependence on the input DMs. Thus it is 
very important to study the default model depedence of the output spectral functions.

 \begin{figure}[f]
  \begin{center}
  \leavevmode
  \includegraphics[width=0.5\textwidth]{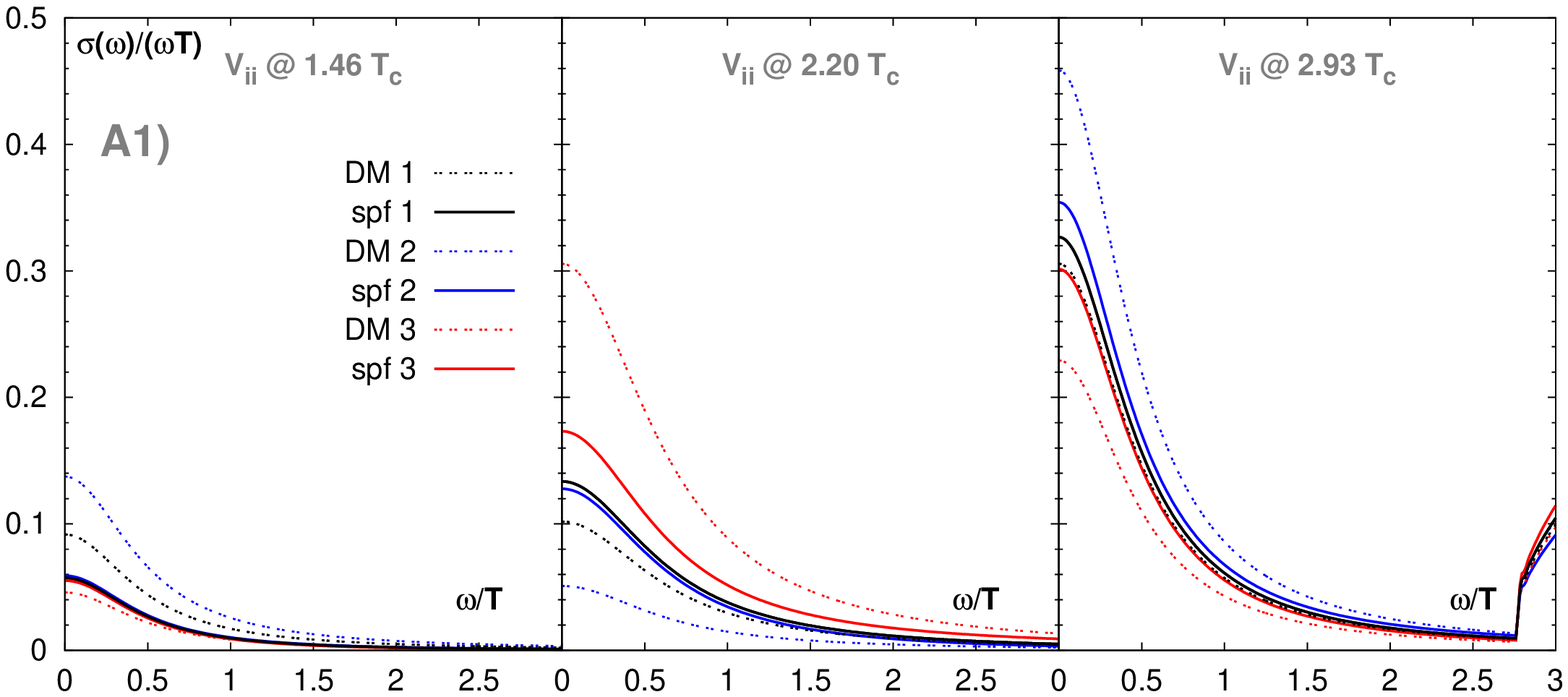}~\includegraphics[width=0.5\textwidth]{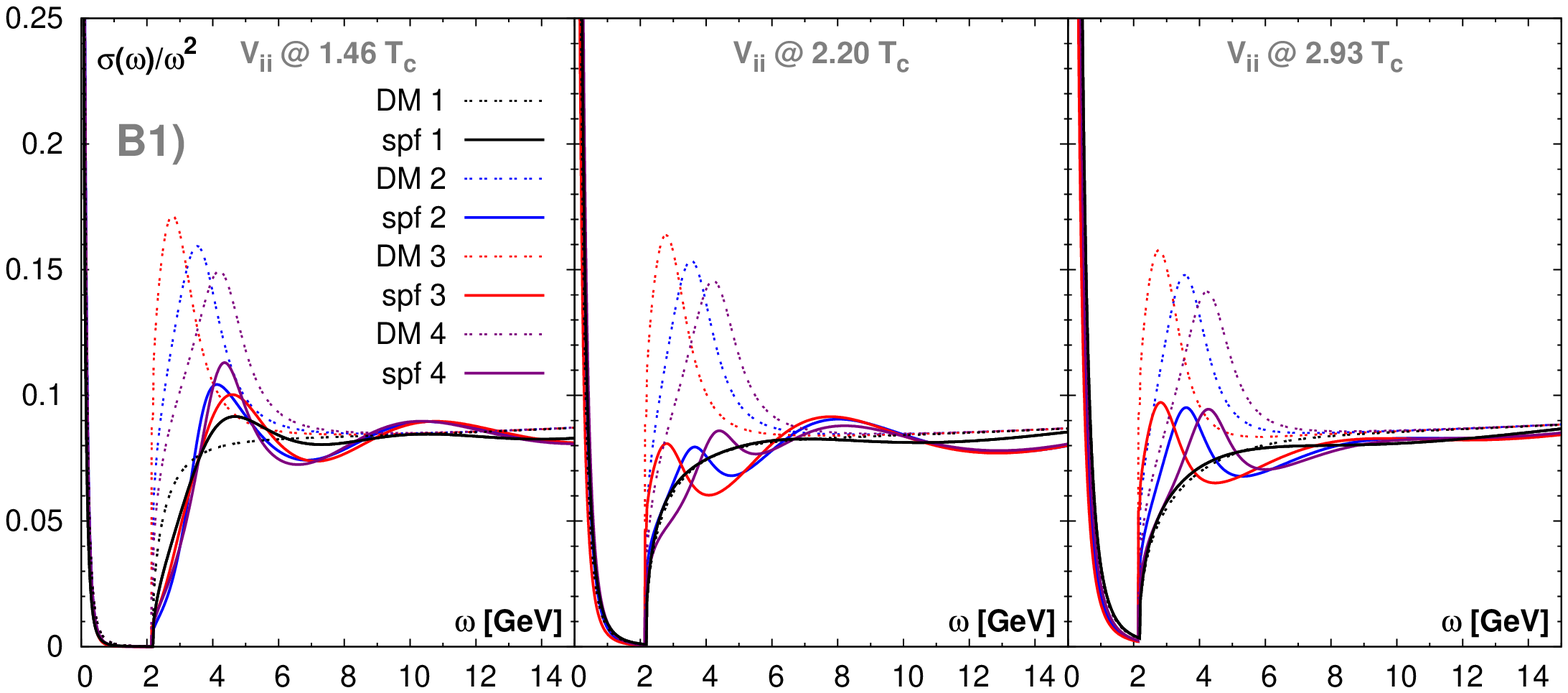}\\
 \includegraphics[width=0.5\textwidth]{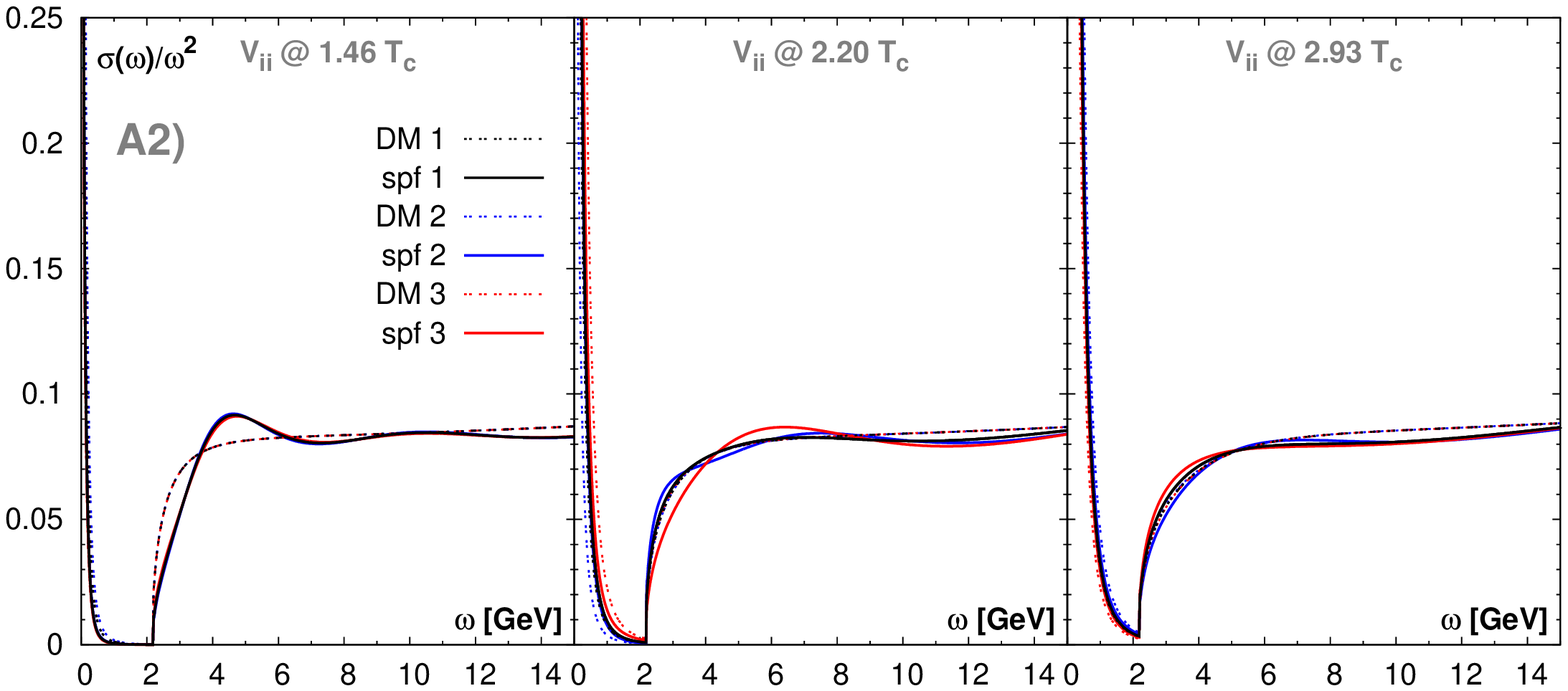}~\includegraphics[width=0.5\textwidth]{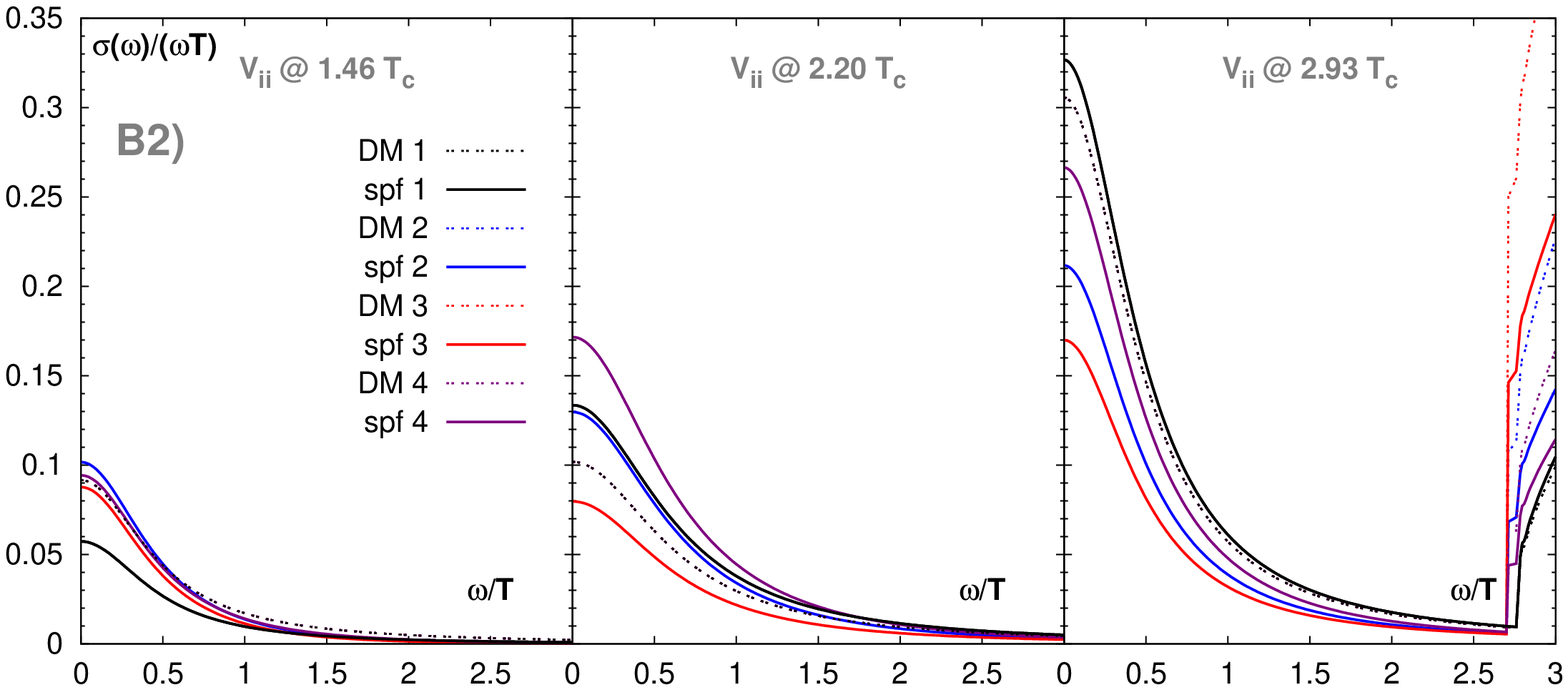}
\end{center}
    \caption{Default model dependences of vector spectral functions at $T>T_c$. The dotted and corresponding solid lines are for the different default models and output spectral functions, respectively.  }
                 \label{fig:spf_dm}
\end{figure}

      \begin{figure}[f]
  \begin{center}
    \includegraphics[width=0.38\textwidth]{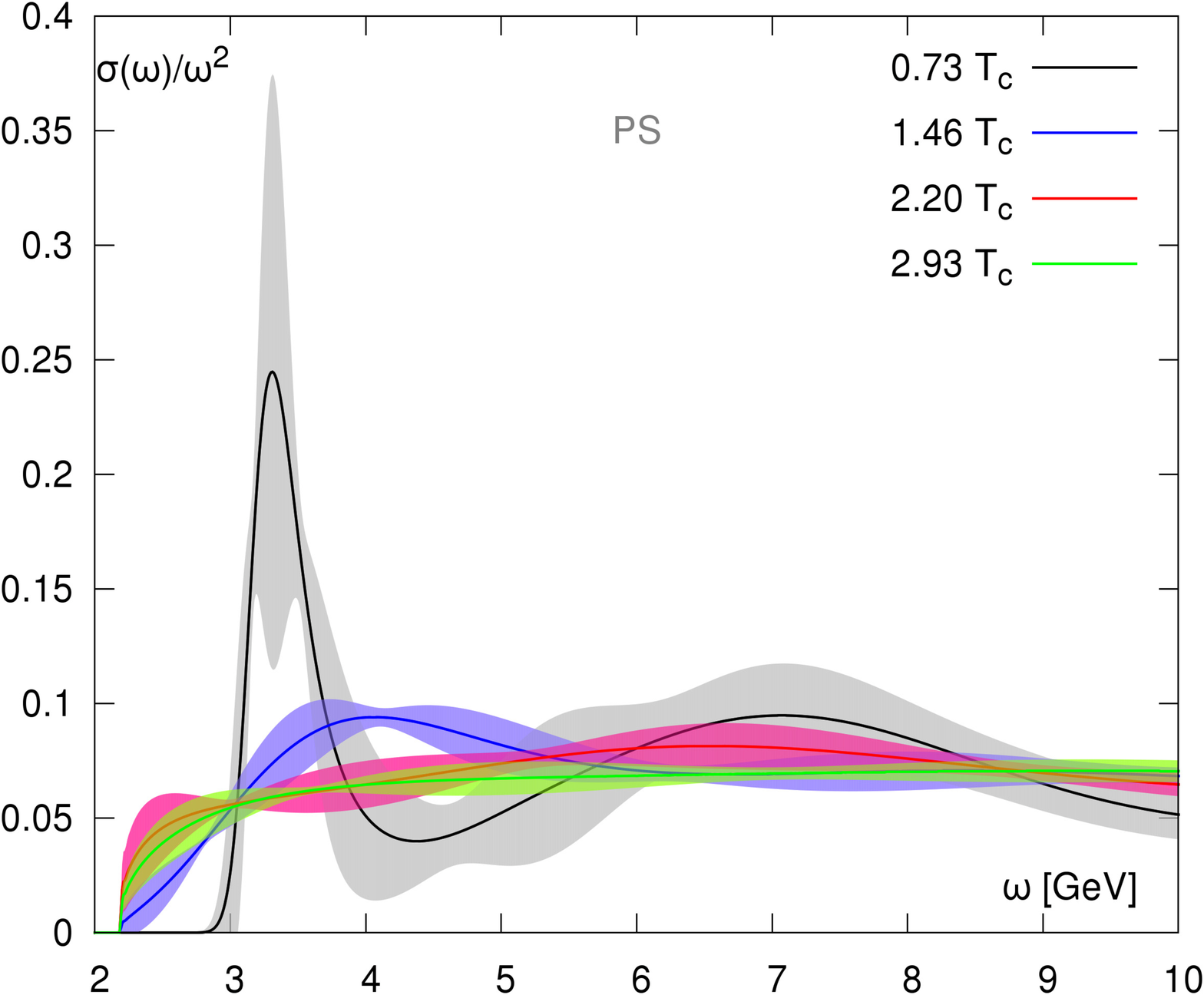}~\includegraphics[width=0.38\textwidth]{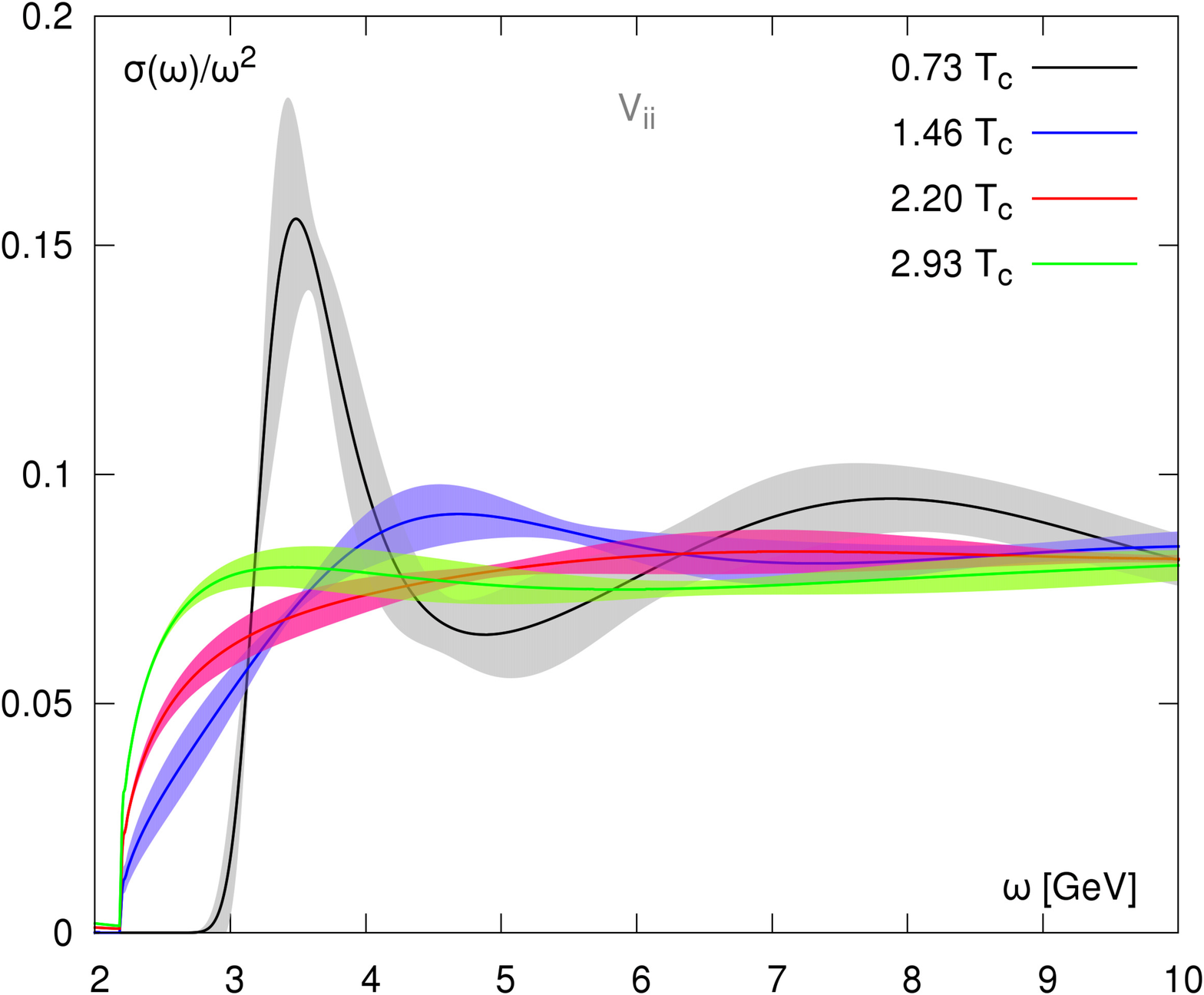}~
                    \caption{Uncertainties of output spectral functions in $PS$  (left) and $V_{ii}$ (right) channels at all available temperatures.  The shaded areas are 
errors of output spectral functions from Jackknife and 
the solid lines inside the shaded areas are mean values of spectral functions.
                              }
                 \label{fig:res}
  \end{center}
\end{figure}
      \begin{figure}[h]
  \begin{center}
  \includegraphics[width=0.38\textwidth]{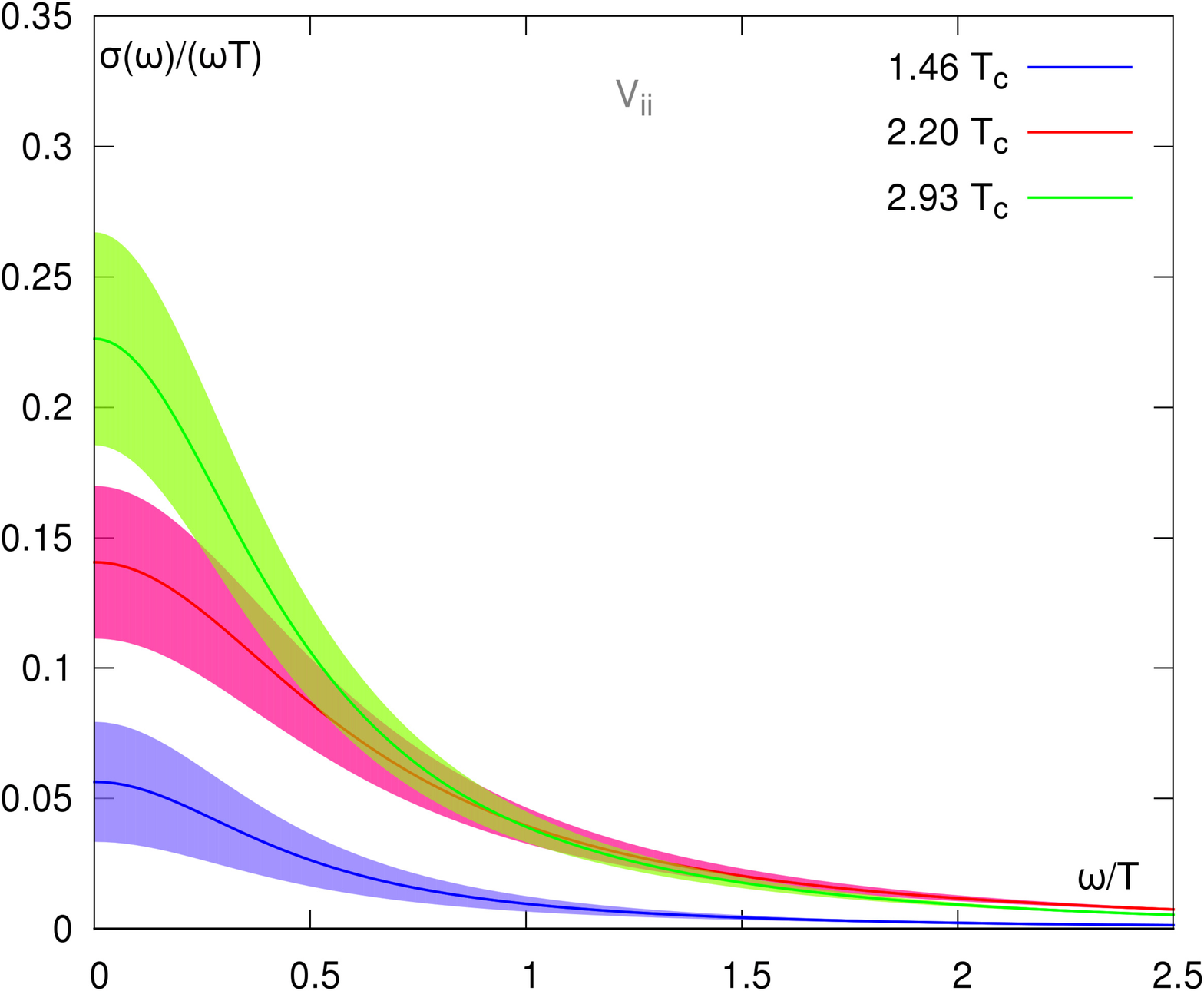}~ \includegraphics[width=0.38\textwidth,height=0.32\textwidth]{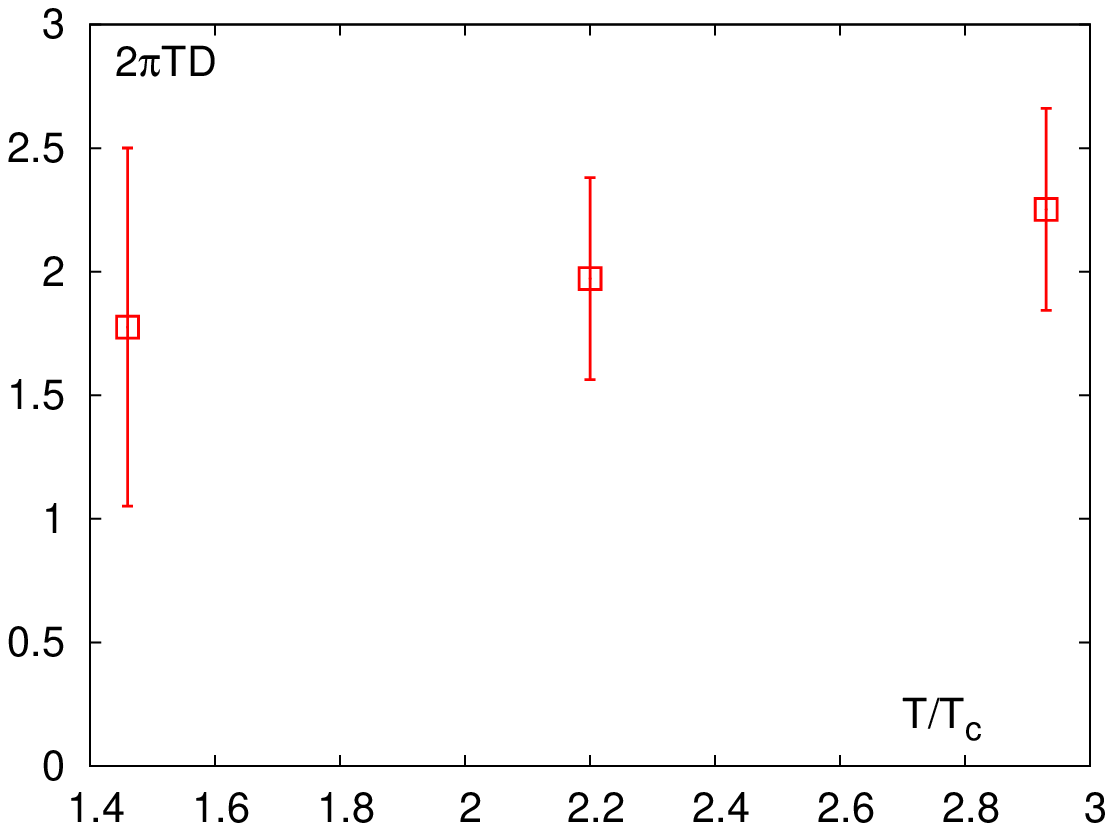}~
                \caption{Transport peaks at $T>T_c$ and resulting diffusion coefficients.
}
                 \label{fig:spf_trans}
\end{center}
\end{figure}

From the left panel of Fig.~\ref{fig:cor} we see that the default model dependence of vector spectral 
functions at $0.73 T_c$ is very small. 
The location of the ground state peak is very close to the 
physical $\Jpsi$ mass. The width of the peak is too wide to be interpreted as the 
physical width of $\Jpsi$. We found no evidence for the existence of a transport peak at this temperature.
In the right panel of Fig.~\ref{fig:cor} we 
show $(G(\tau T)-G_{rec}(\tau T))=\int\md\omega K(\tau T)(\sigma(\omega,T)-\sigma(\omega,0.73T_c))$  as a function of $\tau T$.
The flatness of $G-G_{rec}$ at $1.46 T_c$ indicates some small changes in the bound states and 
the rising feature with increasing distances at higher temperatures 
indicates  $(\sigma(\omega,T)-\sigma(\omega,0.73T_c))$ is negative in some 
low frequency region. To look into the detailed change of the spectral function from below to above $T_c$, it is crucial 
to investigate the spectral function itself.

In Fig.~\ref{fig:spf_dm} we show the default model dependence of vector spectral 
functions at $T>T_c$. We first vary the low 
frequency (transport) part of the spectral function in the default model, where plots A1) 
and A2) show the low 
frequency and high frequency parts of the spectral function, respectively.
A small default model dependence is observed. We then vary the resonance part of the spectral 
function in the default model, where plots B1) and B2) show the high frequency and low frequency parts of the spectral 
function, respectively. In particular, the first peak location in ``DM 2" corresponds 
to the peak location of the spectral function at $0.73T_c$. We observed that the ground 
state peak location in general shifts to higher energy region at $1.46T_c$ and becomes flat at higher 
temperatures. Transport peaks have relatively strong default model dependences as seen in plot B2).

\section{Conclusion}
We summarize current uncertainties of transport peaks in the left panel of Fig.~\ref{fig:spf_trans} and 
resulting heavy quark diffusion coefficients in the right panel. We found that $2\pi TD$ at 1.46 $T_c$ is  close to unity and is slightly 
increasing with temperature. We also performed the default model dependence study in the PS channel and together with
the uncertainties of vector spectral function are shown in Fig.~\ref{fig:res}, which suggests the dissociation 
of both $\Jpsi$ and $\eta_c$ at $T\geqslant1.46~T_c$.


\vspace{-0.5cm}
\section*{Acknowledgments}
This work has been supported in part by the Deutsche 
Forschungsgemeinschaft under grant GRK 881 and by contract DE-AC02-98CH10886 
with the U.S. Department of Energy. HTD thanks G. Moore
for correspondences on NLO pQCD results of heavy quark diffusion.


\section*{References}

\end{document}